\documentclass[nofootinbib,twocolumn,showpacs,preprintnumbers,pre,aps]{revtex4-1}

\usepackage{graphicx}
\usepackage{bm}
\usepackage{amsmath}
\usepackage{amssymb}
\usepackage{color}

\begin{document}

\title{Reciprocal microswimmers in a viscoelastic fluid}%

\author{Kento Yasuda}

\author{Mizuki Kuroda}

\author{Shigeyuki Komura}\email{komura@tmu.ac.jp}

\affiliation{
Department of Chemistry, Graduate School of Science,
Tokyo Metropolitan University, Tokyo 192-0397, Japan}


\begin{abstract}
We suggest several reciprocal swimming mechanisms that lead to a locomotion only in viscoelastic fluids.
In the first situation, we consider a three-sphere microswimmer with a difference in oscillation 
amplitudes for the two arms. 
In the second situation, we consider a three-sphere microswimmer in which one of the frequencies 
of the arm motion is twice as large as the other one.
In the third situation, we consider a two-sphere microswimmer with a difference in size for the 
two spheres.
In all these three cases, the average velocity is proportional to the imaginary part of the 
complex shear viscosity of a surrounding viscoelastic medium.
We show that it is essential for a micromachine to break its structural symmetry in order to swim in a 
viscoelastic fluid by performing reciprocal body motions.
\end{abstract}

\maketitle

\section{Introduction}
\label{sec:introduction}

Microswimmers are small machines that swim in a fluid and have potential applications in 
microfluidics and microsystems~\cite{Lauga09}.
Over the length scale of microswimmers, the fluid forces acting on them are dominated by the frictional 
viscous forces.
By transforming chemical energy into mechanical energy, however, microswimmers change their 
shape and move efficiently in viscous environments.
According to the scallop theorem suggested by Purcell, reciprocal body motion cannot be 
used for locomotion in a Newtonian fluid~\cite{Purcell77,Lauga11,Ishimoto12}.
As one of the simplest models exhibiting non-reciprocal body motion, 
Najafi and Golestanian proposed a three-sphere swimmer~\cite{Golestanian04,Golestanian08}, 
in which three in-line spheres are linked by two arms of varying length.
Recently, such a swimmer has been experimentally realized by using colloidal beads manipulated by 
optical tweezers~\cite{Leoni09}, ferromagnetic particles at an air-water 
interface~\cite{Grosjean16,Grosjean18}, or neutrally buoyant spheres in a viscous 
fluid~\cite{Box17}.

For many microswimmers in nature, however, the surrounding fluid is not necessarily purely viscous 
but in general viscoelastic.
Several studies have discussed the swimming behaviors of micromachines in different types of viscoelastic 
fluids~\cite{Fu07,Fu09,Lauga09b,Teran10,Curtis13,Qiu14,Ishimoto17,Datt18}.
In particular, Lauga showed that the scallop theorem in a viscoelastic fluid breaks down if the 
squirmer has a fore-aft asymmetry in its surface velocity distribution~\cite{Lauga09b}.
In our recent study, we have discussed the locomotion of a three-sphere microswimmer in a 
viscoelastic medium~\cite{Yasuda17a}.
Here a relationship linking the average swimming velocity to the frequency-dependent viscosity 
of the surrounding medium was derived. 
We demonstrated that the absence of the time-reversal symmetry of the body motion 
(i.e., non-reciprocal motion) is reflected in the real part of the frequency-dependent complex 
viscosity, whereas the absence of the structural symmetry of the swimmer shape is reflected 
in its imaginary part~\cite{Yasuda17a}.

Later, we investigated the locomotion of a three-sphere microswimmer in a viscoelastic structured fluid   
characterized by typical length and time scales~\cite{Yasuda18}.
The competition between the swimmer size and the characteristic length scale associated with the 
fluid internal structure gives rise to the rich dynamics~\cite{WittenBook,LarsonBook}.
The present authors have also proposed a generalized three-sphere microswimmer model 
in which the spheres are connected by two harmonic springs, i.e., an elastic 
microswimmer~\cite{Yasuda17b,Hosaka17,Kuroda19,Sou19}.  
It has been shown that an elastic microswimmer in a purely viscous fluid exhibits ``viscoelastic"
effects as a whole~\cite{Yasuda17b,Kuroda19}.

In this paper, employing either a three-sphere or a two-sphere microswimmer, 
we suggest several swimming mechanisms which include only reciprocal (rather than non-reciprocal) 
body motions and can lead to a locomotion only in viscoelastic fluids.
According to the scallop theorem~\cite{Purcell77,Lauga11,Ishimoto12}, the considered reciprocal body 
motions cannot be used for locomotion in a purely viscous fluid.
For a three-sphere swimmer in a viscoelastic fluid, the simplest reciprocal body motion has 
been proposed in our previous work~\cite{Yasuda17a}.
This is possible when the two amplitudes of the oscillatory arm motion are different, namely,  
when the structural symmetry of a three-sphere microswimmer is broken.  
For the illustration of the calculation scheme, we first explain this reciprocal motion even though 
the result is a part of the calculation in Ref.~\cite{Yasuda17a}.

We then suggest two other reciprocal swimming mechanisms in a general viscoelastic fluid; 
a three-sphere microswimmer in which one of the frequencies of the arm motion is twice as 
large as the other one, and a two-sphere microswimmer with a difference in size for the 
two spheres.
In all these three cases, we show that the average velocity is proportional to the imaginary part of 
the complex shear viscosity that characterizes the elasticity of the surrounding fluid. 
The suggested body motions highlight the essential swimming mechanism of a micromachine
in viscoelastic fluids.  
For the sake of clarity, we do not include any non-reciprocal body motions of a microswimmer
as discussed in Ref.~\cite{Yasuda17a}. 
Moreover, we assume that the surrounding viscoelastic fluid is homogeneous and do not consider 
any fluid internal structures as in Ref.~\cite{Yasuda18}.

\begin{figure}[htb]
\begin{center}
\includegraphics[scale=0.4]{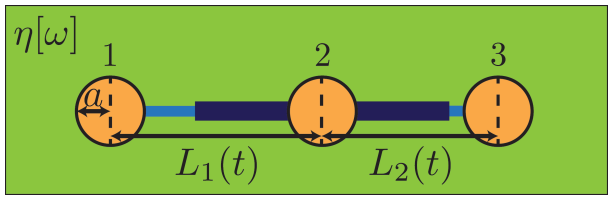}
\end{center}
\caption{
Najafi--Golestanian three-sphere swimmer model.
Three identical spheres of radius $a$ are connected by arms of lengths $L_1(t)$ and  $L_2(t)$, 
and they undergo time-dependent cyclic motions
(see Eqs.~(\ref{L1cos}) and (\ref{L2cos}) or Eqs.~(\ref{VEeq:larmlength}) and (\ref{VEeq:armlength})).
Such a microswimmer is embedded in a viscoelastic medium characterized by a frequency-dependent 
complex shear viscosity $\eta[\omega]$. 
In this work, we consider only reciprocal body motions. 
}
\label{model}
\end{figure}

In the next section, we briefly review Ref.~\cite{Yasuda17a} to show the basic 
equations for the motion of a three-sphere swimmer in a general viscoelastic fluid. 
In Sec.~\ref{sec:amplitude}, we discuss the locomotion of a three-sphere swimmer when 
the two arm amplitudes are asymmetric, as already discussed in Ref.~\cite{Yasuda17a}.
In Sec.~\ref{sec:frequency}, we explain the case of asymmetric arm frequencies for a 
three-sphere swimmer.
The generalization for higher frequencies of the arm motion is also discussed. 
In Sec.~\ref{sec:two-sphere}, we present the result for an asymmetric two-sphere 
microswimmer in a viscoelastic fluid.
Finally, a summary of our work and a discussion is provided in Sec.~\ref{sec:discussion}.

\section{Three-sphere microswimmer in a viscoelastic fluid}
\label{sec:three-sphere}

The general equation that describes the hydrodynamics of a low-Reynolds-number flow in  
a viscoelastic medium is given by the following generalized Stokes equation~\cite{Granek11}:
\begin{equation}
\int_{-\infty}^t dt'\,\eta(t-t') \nabla^2 \mathbf{v}(\mathbf{r},t') -\nabla p(\mathbf{r},t)=0.
\label{StorksEve}
\end{equation}
Here $\eta(t)$ is the time-dependent shear viscosity, $\mathbf{v}$ is the velocity field, 
$p$ is the pressure field, and $\mathbf{r}$ stands for a three-dimensional positional vector.
The above equation is further subjected to the incompressibility condition, 
\begin{equation}
\nabla\cdot\mathbf{v}=0.
\end{equation}
From these equations, one can obtain a linear relation between the 
time-dependent force $F(t)$ acting on a hard sphere of radius $a$ and its 
time-dependent velocity $V(t)$.
In the Fourier domain, this relation can be represented as 
\begin{equation}
V(\omega)=\frac{1}{6\pi\eta[\omega] a}F(\omega),
\label{GSR}
\end{equation}
where we use a bilateral Fourier transform for 
$V(\omega)=\int_{-\infty}^{\infty} dt\, V(t)e^{-i\omega t}$ and 
$F(\omega)=\int_{-\infty}^{\infty} dt\, F(t)e^{-i\omega t}$, 
while we employ a unilateral one for 
$\eta[\omega]=\int_{0}^{\infty} dt\, \eta(t)e^{-i\omega t}$.
Equation~(\ref{GSR}) is the generalized Stokes-Einstein relation (GSR), which has been 
successfully used in active microrheology experiments~\cite{GSOMS,Schnurr97,Chen10},
and its mathematical validity has also been discussed~\cite{SM10,FurstBook}.

Next, we briefly explain the three-sphere micromachine model proposed by Najafi and 
Golestanian~\cite{Golestanian04,Golestanian08}. 
As schematically shown in Fig.~\ref{model}, this model consists of three spheres of 
the same radius $a$.
They are connected by two arms of lengths $L_1(t)$ and $L_2(t)$, which undergo 
time-dependent motion, as we will discuss separately in the next sections.
Moreover, the radius of the two arms is assumed to be negligibly small.
If we define the velocity of each sphere along the swimmer axis as $V_i(t)$ 
($i=1,2,3$), we have 
\begin{align}
\dot{L}_1(t)&=V_2(t)-V_1(t),
\label{L1dot}
\\
\dot{L}_2(t)&=V_3(t)-V_2(t),
\label{L2dot}
\end{align}
where $\dot{L}_1$ and $\dot{L}_2$ indicate the time derivatives of $L_1$ and $L_2$, 
respectively.

Owing to the hydrodynamic effect, each sphere exerts a force $F_i$  on the viscoelastic 
medium and experiences a force $-F_i$ from it.
To relate the forces and the velocities in the frequency domain, we use the GSR in 
Eq.~(\ref{GSR}) and the Oseen tensor, in which the frequency-dependent viscosity 
$\eta[\omega]$ is used instead of a constant one~\cite{MW95,Mason00}.
Assuming that $a\ll L_1, L_2 $, we can write the three velocities $V_i(\omega)$ 
as~\cite{Golestanian04,Golestanian08} 
\begin{widetext}
\begin{align}
\label{eq3}
V_1(\omega)&=\frac{F_1(\omega)}{6\pi\eta[\omega] a}+\frac{1}{4\pi\eta[\omega]}\frac{F_2(\omega) \ast L_1^{-1}(\omega)}{2\pi}+\frac{1}{4\pi\eta[\omega]}\frac{F_3(\omega) \ast (L_1+L_2)^{-1}(\omega)}{2\pi},\\
\label{eq4}
V_2(\omega)&=\frac{1}{4\pi\eta[\omega]} \frac{F_1(\omega) \ast L_1^{-1}(\omega)}{2\pi}+\frac{F_2(\omega)}{6\pi\eta[\omega] a}+\frac{1}{4\pi\eta[\omega] }\frac{F_3(\omega) \ast L_2^{-1}(\omega)}{2\pi},\\
\label{eq5}
V_3(\omega)&=\frac{1}{4\pi\eta[\omega] }\frac{F_1(\omega)\ast (L_1+L_2)^{-1}(\omega)}{2\pi}+\frac{1}{4\pi\eta[\omega] }\frac{F_2(\omega) \ast L_2^{-1}(\omega)}{2\pi}+\frac{F_3(\omega)}{6\pi\eta[\omega] a},
\end{align}
\end{widetext}
where we have used bilateral Fourier transforms such as 
$L_1^{-1}(\omega)=\int_{-\infty}^\infty dt\,[L_1(t)]^{-1}e^{-i\omega t}$.
Furthermore, the convolution of two functions is generally defined by 
$g_1(\omega)\ast g_2(\omega)=\int _{-\infty}^\infty d\omega'\, g_1(\omega-\omega')g_2(\omega')$
in the above equations.

Since we are interested in the autonomous net locomotion of the swimmer, 
there are no external forces acting on the spheres.
Neglecting the inertia of the surrounding fluid, we require the following force balance condition: 
\begin{equation}
F_1(\omega)+F_2(\omega)+F_3(\omega)=0.
\label{eq6}
\end{equation}

Since Eqs.~(\ref{eq3})--(\ref{eq5}) involve convolutions in the frequency domain, 
we cannot solve these equations for arbitrary $L_1(t)$ and $L_2(t)$. 
In the subsequent sections, we assume three different reciprocal arm motions for $L_{1}(t)$ and $L_{2}(t)$,
and obtain the average velocity of a microswimmer in a viscoelastic fluid.

\section{Asymmetric arm amplitudes}
\label{sec:amplitude}

We first consider the case when the amplitudes of the two arms are different. 
We assume that the two arms undergo the following reciprocal periodic motion: 
\begin{align}
L_1(t)&=\ell+d_1\cos(\Omega t),
\label{L1cos}
\\
L_2(t)&=\ell+d_2\cos(\Omega t).
\label{L2cos}
\end{align}
In the above, $\ell$ is the constant length, $d_1$ and $d_2$ are the amplitudes of the oscillatory 
motion, $\Omega$ is the common arm frequency.
It should be emphasized that, in contrast to Ref.~\cite{Yasuda17a}, we do not include any difference 
in the phases between the two arms, and hence the whole body motion is reciprocal.
On the other hand, we characterize the structural symmetry of the swimmer by $d_1$ and $d_2$.
The whole micromachine is symmetric when $d_1=d_2$, while it is asymmetric when $d_1 \ne d_2$.

Since the arm frequency is $\Omega$,  we assume that the velocities 
and the forces of the three spheres can generally be written as 
\begin{align}
V_i(\omega)& =V_{i,0}\,\delta(\omega)\nonumber\\
&+\sum_{n=1}^\infty \left[V_{i,n}\,\delta(\omega+n\Omega)+V_{i,-n}\,\delta(\omega-n\Omega)\right],
\label{Vex}\\
F_i(\omega)& =F_{i,0}\,\delta(\omega)\nonumber\\
&+\sum_{n=1}^\infty \left[F_{i,n}\,\delta(\omega+n\Omega)+F_{i,-n}\,\delta(\omega-n\Omega)\right].
\label{Fex}
\end{align}
Substituting Eqs.~(\ref{Vex}) and (\ref{Fex}) into the six coupled 
Eqs.~(\ref{L1dot})--(\ref{eq6}),  we obtain in general a matrix equation with infinite dimensions.

\begin{figure}[thb]
\begin{center}
\includegraphics[scale=0.5]{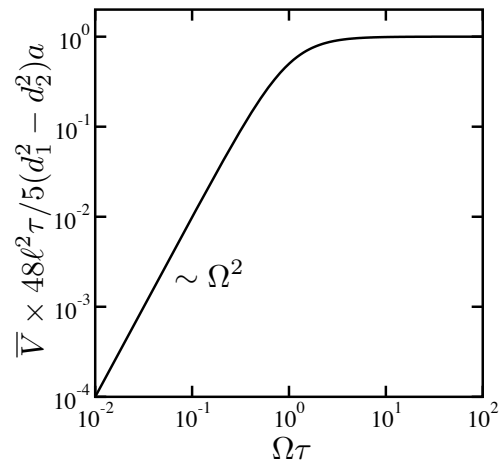}
\end{center}
\caption{
Average swimming velocity $\overline{V}$ as a function of $\Omega \tau$, 
where $\Omega$ is the arm frequency and $\tau$ is the characteristic time 
scale in the Maxwell model.
Here $\overline{V}$ is scaled by $5(d_1^2-d_2^2)a/(48\ell^2\tau)$ assuming that 
$d_{1} \neq d_{2}$.
$\overline{V}$ increases as $\overline{V} \sim \Omega^2$ for $\Omega \tau \ll 1$.}
\label{plotofV}
\end{figure}

Under the conditions $d_1, d_2 \ll \ell$ and $a \ll \ell$,  we are allowed to consider only 
$n=0$, $\pm 1$, and we further use the approximation $F_{i,\pm2}\approx 0$.
Then we can solve for the six unknown functions $V_i(\omega)$ and $F_i(\omega)$, 
and also calculate the total swimming velocity 
\begin{equation}
V=\frac{1}{3}(V_1+V_2+V_3). 
\end{equation}
Up to the lowest order terms in $a$, the average swimming velocity over one 
cycle of motion becomes~\cite{Yasuda17a}
\begin{equation}
\overline{V} =  
-\frac{5a(d_1^2-d_2^2)\Omega}{48\ell^2\eta_0}\eta''[\Omega],
\label{barVve}
\end{equation}
where $\eta''[\Omega]$ is the imaginary part of the complex shear viscosity, 
$\eta[\Omega]=\eta'[\Omega]+i\eta''[\Omega]$, and $\eta_0=\eta[\Omega \rightarrow 0]$ 
is the constant zero-frequency viscosity.
A detailed derivation of Eq.~(\ref{barVve}) is given in the Appendix A.
Notice that $\eta''[\Omega]$ is taken to be negative in our notation. 
Hence $\overline{V}>0$ when $d_{1}>d_{2}$.

Since Eq.~(\ref{barVve}) involves $\eta''[\Omega]$, it can be regarded as an elastic contribution 
that exists when the structural symmetry of the swimmer is broken, i.e., $d_1 \ne d_2$. 
In other words, a reciprocal three-sphere micromachine uses the elastic degree of freedom 
of the surrounding viscoelastic medium for its locomotion. 
The structural asymmetry, $d_1 \ne d_2$, is necessary for a microswimmer to determine 
its moving direction.
For a purely Newtonian fluid, namely, for a medium characterized by a constant viscosity, 
Eq.~(\ref{barVve}) vanishes even when $d_1 \ne d_2$ because $\eta''[\Omega]=0$.
The above result also implies that a three-sphere swimmer cannot move in a purely elastic 
medium, for which we have $\eta_0 \rightarrow \infty$.

When the arm motion is non-reciprocal, such as by introducing a phase difference between 
the two arms, a different term arises~\cite{Yasuda17a,Yasuda18}.
This term includes $\eta'[\Omega]$ and hence can be regarded as the viscous contribution.  
Because Eq.~(\ref{barVve}) contributes to the average velocity even for a reciprocal body 
motion, the scallop theorem should be generalized for a three-sphere 
swimmer in a viscoelastic medium~\cite{Lauga09b}.

To illustrate the above result, we assume that the surrounding viscoelastic medium 
is described by a simple Maxwell model~\cite{Yasuda17a}.  
In this case, the frequency-dependent complex viscosity can be written as 
\begin{equation}
\eta[\omega]=\eta_0\frac{1-i\omega\tau}{1+\omega^2\tau^2}, 
\label{maxwell}
\end{equation}
where $\tau$ is the characteristic time scale.
Within this model, the medium behaves as a viscous fluid for $\omega\tau \ll 1$, 
while it becomes elastic for $\omega\tau \gg 1$.
Using Eq.~(\ref{maxwell}), we can easily obtain the average swimming velocity in 
Eq.~(\ref{barVve}) as~\cite{Yasuda17a} 
\begin{align}
\overline{V} = \frac{5(d_1^2-d_2^2)a\Omega}{48\ell^2}
\frac{\Omega\tau}{1+\Omega^2\tau^2}.
\label{maxwellV}
\end{align}
Here $\overline{V}$ increases as $\overline{V} \sim \Omega^2$ for 
$\Omega \tau \ll 1$, and it approaches a constant for $\Omega \tau \gg 1$.
In Fig.~\ref{plotofV}, we plot the dimensionless average swimming velocity $\overline{V}$ as 
a function of the dimensionless arm frequency $\Omega \tau$ when $d_1 \ne d_2$.

\section{Asymmetric arm frequencies}
\label{sec:frequency}

As the second case, we consider the situation where the frequencies of the two arms are different.
For the sake of simplicity, we consider here the following time dependencies:
\begin{align}
L_{1}(t)&=\ell+d\cos(\Omega t),
\label{VEeq:larmlength}\\
L_{2}(t)&=\ell+d\cos(2\Omega t).
\label{VEeq:armlength}
\end{align}
In the above, the frequency of $L_{2}$ is twice as large as that of $L_{1}$, whereas the
amplitude of oscillation $d$ is taken to be the same. 
Since the arm frequencies are different, a phase shift does not play any role, and 
the overall arm motion can be regarded as reciprocal for Eqs.~(\ref{VEeq:larmlength}) and 
(\ref{VEeq:armlength}).

The procedure to obtain the average velocity is essentially the same as in the previous section. 
We assume that the velocities and the forces of the three spheres are also expressed by 
Eqs.~(\ref{Vex}) and (\ref{Fex}).
Under the conditions $d\ll\ell$ and $a\ll\ell$, we consider only $n=0$, $\pm1$, $\pm2$ and 
use the approximation $F_{i,\pm 3}\approx0$ because of Eq.~(\ref{VEeq:armlength}).
After some calculation, the average swimming velocity can be obtained as 
\begin{align}
\overline{V} = -\frac{5a d^2 \Omega}{48 \ell^2 \eta_0}
\left( \eta''[\Omega]-2 \eta''[2\Omega] \right).
\label{VEeq:res1}
\end{align}

Similar to Eq.~(\ref{barVve}), only the imaginary part of the complex shear viscosity appears 
in the above expression, and the two terms in Eq.~(\ref{VEeq:res1}) are the elastic contributions. 
The above result means that a micromachine can swim as long as $\eta''[\Omega]\neq 2\eta''[2\Omega]$ 
which usually holds for viscoelastic fluids.
It is interesting to note that the direction of locomotion is determined by the relative magnitude 
between $\eta''[\Omega]$ and $2\eta''[2\Omega]$.
When the arm amplitudes are different and characterized by $d_1$ and $d_2$, as in Eqs.~(\ref{L1cos}) 
and (\ref{L2cos}), we have confirmed that the average velocity is then proportional to 
$d_{1}^{2}\eta''[\Omega] - 2d_{2}^{2}\eta''[2\Omega]$, as one can expected from 
Eqs.~(\ref{barVve}) and (\ref{VEeq:res1}).

In general, the motions of the two arms can be given by  
\begin{align}
L_{1}(t)&=\ell+d\cos(\Omega t),
\label{VEeq:larmlength2}\\
L_{2}(t)&=\ell+d\cos(m\Omega t),
\label{VEeq:armlength2}
\end{align} 
where $m$ is an integer.
Notice that the average velocity vanishes for $m=1$ even in a viscoelastic fluid because the 
arm amplitudes are the same in Eqs.~(\ref{VEeq:larmlength2}) and (\ref{VEeq:armlength2}).
Although we have explicitly calculated only up to $m=3$, we speculate that the average velocity 
can be given by 
\begin{align}
\overline{V} = -\frac{5ad^{2}\Omega}{48\ell^2\eta_0}
\left(\eta''[\Omega]-m\eta''[m\Omega] \right),
\label{VEeq:predi}
\end{align}
which is a natural generalization of Eq.~(\ref{VEeq:res1}).
When $m$ is very large, the first term becomes negligible, and the whole locomotion is 
dominated by $\eta''[m\Omega] $.

One can further generalize Eq.~(\ref{VEeq:larmlength2})  to $L_{1}(t)=\ell+d\cos(M \Omega t)$,
where $M$ is another integer, while $L_{2}$ is still given by Eq.~(\ref{VEeq:armlength2}) but 
$M \neq m$.  
Then the least common multiple of $M$ and $m$ determines the period of the overall reciprocal 
motion of a micromachine.
In this case,  we predict in general that the first term in Eq.~(\ref{VEeq:predi}) will be replaced by 
$M \eta''[M \Omega]$ which results from the symmetry of our system.

\section{Asymmetric two-sphere microswimmer}
\label{sec:two-sphere}

\begin{figure}[t]
\begin{center}
\begin{center}
\includegraphics[scale=0.5]{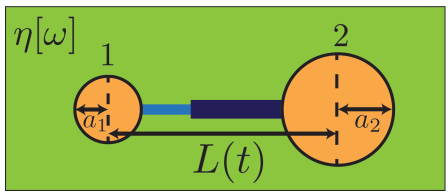}
\end{center}
\caption{Asymmetric two-sphere swimmer model. 
Two spheres of different radius $a_1$ and $a_2$ ($a_1<a_2$) are connected by an arm of length 
$L(t)$, and it undergoes a time-dependent periodic motion (see Eq.~(\ref{VEeq:twoswimarmlength})).
The swimmer is embedded in a viscoelastic medium characterized by a frequency-dependent complex 
shear viscosity $\eta[\omega]$.} 
\label{twosphereswimmerpic}
\end{center}
\end{figure}

As the third reciprocal body motion, we consider a two-sphere swimmer consisting of two 
hard spheres having different sizes.
As shown in Fig.~\ref{twosphereswimmerpic}, these two spheres are connected by a single 
arm which can vary its length.
The radii of the two spheres are denoted by $a_1$ and $a_2$, and the distance between them is $L(t)$.
As the equations of motion for the two spheres are even simpler than those for a three-sphere 
swimmer, we shall explicitly write them below.

Similar to Eqs.~(\ref{L1dot}) and (\ref{L2dot}), the time derivative of $L$ is given by 
\begin{align}
\dot{L}(t)=V_{2}(t)-V_{1}(t).
\label{VEeq:fouri3}
\end{align}
Corresponding to Eqs.~(\ref{eq3})--(\ref{eq5}), the relations between the velocities and the forces
in the frequency domain can be written as 
\begin{align}
V_{1}(\omega)  & =  \frac{F_1(\omega)}{6\pi\eta[\omega] a_{1}}
+\frac{1}{4\pi\eta[\omega]}\frac{F_{2}(\omega) \ast L^{-1}(\omega)}{2\pi},
\label{VEeq:twoveleq1}\\
V_{2}(\omega)  & = \frac{1}{4\pi\eta[\omega]}\frac{F_{1}(\omega) \ast L^{-1}(\omega)}{2\pi}
+\frac{F_2(\omega)}{6\pi\eta[\omega] a_{2}}.
\label{VWeq:twoveleq2}
\end{align}
Finally, the force balance equation now becomes 
\begin{align}
F_{1}(\omega)+F_{2}(\omega)=0.
\label{VEeq:twoswimvalance}
\end{align}

The periodic arm motion is assumed to have the following simple form: 
\begin{align}
L(t)=\ell+d\cos{(\Omega t)}.
\label{VEeq:twoswimarmlength}
\end{align}
Since there is only one arm, it is obvious that any periodic arm motion is inevitably reciprocal.
Under the conditions $d \ll \ell$ and $a_1, a_2 \ll \ell$,  we consider only $n=0$, $\pm 1$ and 
use the approximation $F_{i,\pm2}\approx 0$ in Eqs.~(\ref{Vex}) and~(\ref{Fex}).
Calculating the total swimming velocity $V=(V_1+V_2)/2$, we finally obtain the average swimming 
velocity over one cycle of motion as 
\begin{align}
\overline{V} = \frac{3 a_{1}a_{2}(a_{1}-a_{2})d^{2}\Omega}{4\ell^2(a_{1}+a_{2})^{2}\eta_0}
\eta''[\Omega].
\label{VEeq:res4}
\end{align}

This result shows that a reciprocal two-sphere micromachine can swim in a viscoelastic fluid when the 
sphere sizes are different, i.e., $a_{1}\neq a_{2}$.
Similar to the previous cases, the average velocity depends only on $\eta''[\Omega]$ and it 
is due to the elastic contribution. 
Hence the elasticity of a viscoelastic medium is responsible for the locomotion of a reciprocal 
microswimmer as long as its structure is asymmetric. 
This statement does not contradict with the original scallop theorem which holds only for purely 
viscous fluids~\cite{Purcell77,Lauga11,Ishimoto12}. 
When the surrounding fluid is purely elastic, however, the average velocity $\overline{V}$ vanishes 
because $\eta_0\to\infty$.

In the limit of $a_1\ll a_2$, for example, Eq.~(\ref{VEeq:res4}) further reduces to 
\begin{align}
\overline{V}\approx - \frac{3a_{1}d^{2}\Omega }{4\ell^2\eta_0}\eta''[\Omega].
\label{VEeq:res5}
\end{align}
This result shows that the average velocity of a two-sphere swimmer is proportional to the 
radius of the smaller sphere, $a_1$.
Since $\eta''[\Omega]<0$ by definition, $\overline{V}>0$ in the limit of Eq.~(\ref{VEeq:res5}).

Here we discuss the connection between a three-sphere microswimmer and a two-sphere 
microswimmer considered in Sections \ref{sec:frequency} and \ref{sec:two-sphere}, respectively.
According to the average velocity in Eq.~(\ref{VEeq:predi}) for a three-sphere microswimmer, 
its locomotion is dominated by $\eta''[m\Omega] $ when $m \gg 1$.
In such a situation, the motion of the first arm $L_1$ appears to be stagnant when compared 
with that of the second arm $L_2$.
Notice that the limiting expression of Eq.~(\ref{VEeq:predi}) for $m \gg 1$ is similar to the 
average velocity in Eq.~(\ref{VEeq:res5}) for a highly asymmetric two-sphere microswimmer, 
i.e., $a_1\ll a_2$.
Although the numerical factors are different between these two limiting expressions,  
their dependence on the structural and dynamical parameters is identical.
Such a similarity between a three-sphere microswimmer and a two-sphere microswimmer
is an interesting feature of reciprocal micromachines in a viscoelastic fluid.

\section{Summary and discussion}
\label{sec:discussion}

In this paper, employing either a three-sphere or a two-sphere microswimmer, we have suggested 
three reciprocal swimming mechanisms that can lead to a locomotion only in viscoelastic fluids.
In the first situation, we consider a three-sphere microswimmer with a difference in oscillation 
amplitudes for the two arms~\cite{Yasuda17a}. 
In the second situation, we consider a three-sphere microswimmer in which one of the frequencies 
of the arm motion is twice as large as the other one.
In the third situation, we consider a two-sphere microswimmer with a difference in size for the 
two spheres.
In all these three cases, the average velocity is proportional to the imaginary part of the 
complex shear viscosity which characterizes the elastic property of the surrounding viscoelastic fluid.  
Hence it is essential for a micromachine to break its structural symmetry in order to swim 
in viscoelastic fluids by performing reciprocal body motions.
Our result also indicates that the scallop theorem should be generalized for microswimmers 
in a viscoelastic fluid.

Lauga considered an axisymmetric squirming motion of a spherical squirmer 
embedded in an Oldroyd-B fluid, which represents a typical polymeric fluid~\cite{Lauga09b}. 
It was reported that the scallop theorem in a viscoelastic fluid breaks down if the squirmer
has fore-aft asymmetry in its surface velocity distribution, which is in accordance with our result.
On the other hand, Curtis and Gaffney showed that the swimming velocity in a viscoelastic 
medium is the same as that in a Newtonian fluid~\cite{Curtis13}.
Recently, the motion of a two-sphere swimmers in viscoelastic fluids has been discussed 
by Datt \textit{et al.}~\cite{Datt18}.
However, their calculations are limited to an Oldroyd-B fluid.
Our treatment using the GSR in Eq.~(\ref{GSR}) is more general because we do not specify 
any frequency dependence of the complex shear viscosity.
We emphasize that our theory applies for all types of linear viscoelastic fluids.

The scallop theorem states that a microswimmer cannot gain any net displacement 
after one cycle of reciprocal body motion when the surrounding fluid is purely 
viscous~\cite{Purcell77,Lauga11}. 
It should be noted that this theorem is correct only when the Reynolds number 
strictly vanishes~\cite{Ishimoto12}.
Lauga showed that oscillatory reciprocal forcing of a solid body leads to net 
translational motion when the Reynolds number is nonzero even when the fluid is 
purely viscous~\cite{Lauga07}.
It was further predicted that the scallop theorem breaks down with inertia in a continuous 
manner as long as there are some spatial broken symmetries which govern the direction 
of the net motion.
In the future, it would be interesting to see the effects of inertia for a reciprocal microswimmer 
in a viscoelastic fluid and to elucidate how the scallop theorem needs to be extended in 
more general situations.

Even though the argument in this work is restricted to an artificial microswimmer, 
we expect that the basic concept can be applied to more complex biological processes 
such as the motion of bacteria, flagellated cellular swimming, and the beating of cilia. 
Since most of these phenomena take place in a viscoelastic environment, we hope that the
suggested mechanisms in this paper will be applicable for more complex biological swimming
objects.

\section*{Data Availability Statements}

The derivation of the results of this study are available from the corresponding author upon reasonable request.

\acknowledgements

We thank T.\ Kato and Y.\ Hosaka for useful discussions.
We also thank S.\ Al-Izzi for his critical reading of the manuscript.
K.Y.\ acknowledges support by a Grant-in-Aid for JSPS Fellows (Grant No.\ 18J21231) from the Japan Society 
for the Promotion of Science (JSPS). 
S.K.\ acknowledges support by a Grant-in-Aid for Scientific Research (C) (Grant No.\ 18K03567 and
Grant No.\ 19K03765) from the JSPS, 
and support by a Grant-in-Aid for Scientific Research on Innovative Areas
``Information Physics of Living Matters'' (Grant No.\ 20H05538) from the Ministry of Education, Culture, 
Sports, Science and Technology of Japan.

\begin{widetext}
\appendix
\section{Derivation of Eq.~(\ref{barVve})}

In this appendix, we show the detailed derivation of Eq.~(\ref{barVve}).
Substituting Eqs.~(\ref{L1cos}) and (\ref{Vex}) into Eq.~(\ref{L1dot}), we obtain 
\begin{align}
&V_{2,0}-V_{1,0}=0, \\
&V_{2,1}-V_{1,1}=-i\pi d_1\Omega, \\
&V_{2,-1}-V_{1,-1}=i\pi d_1\Omega, \\
&V_{2,n}-V_{1,n}=0~~~\text{for $|n| \ge 2$}.
\label{eq1veT}
\end{align}
Similarly, substituting Eqs.~(\ref{L2cos}) and (\ref{Vex}) into Eq.~(\ref{L2dot}), we obtain
\begin{align}
&V_{3,0}-V_{2,0}=0, \\
&V_{3,1}-V_{2,1}=-i\pi d_2\Omega, \\
&V_{3,-1}-V_{2,-1}=i\pi d_2\Omega, \\
&V_{3,n}-V_{2,n}=0~~~\text{for $|n| \ge 2$}. 
\label{eq2veT}
\end{align}

Next we expand Eqs.~(\ref{eq3}), (\ref{eq4}) and (\ref{eq5}) in terms of the small quantities 
$d_1/\ell$ and $d_2/\ell$ while keeping only the lowest order terms. 
Substituting Eqs.~(\ref{Vex}) and (\ref{Fex}) into these three equations, we obtain
\begin{align}
V_{1,n}& \approx \frac{F_{1,n}}{6\pi\eta[-n\Omega] a}+\frac{1}{4\pi\eta[-n\Omega]\ell}\left(F_{2,n}-\frac{d_1F_{2,n+1}}{2\ell}-\frac{d_1F_{2,n-1}}{2\ell}\right)\nonumber\\
&+\frac{1}{4\pi\eta[-n\Omega]\ell}\left(\frac{F_{3,n}}{2}-\frac{d_1F_{3,n+1}}{8\ell}-\frac{d_1F_{3,n-1}}{8\ell}-\frac{d_2F_{3,n+1}}{8\ell}-\frac{d_2F_{3,n-1}}{8\ell}\right),
\label{eq3veT}\\
V_{2,n}& \approx \frac{1}{4\pi\eta[-n\Omega]\ell}\left(F_{1,n}-\frac{d_1F_{1,n+1}}{2\ell}-\frac{d_1F_{1,n-1}}{2\ell}\right)+\frac{F_{2,n}}{6\pi\eta[-n\Omega] a}\nonumber\\
&+\frac{1}{4\pi\eta[-n\Omega] \ell}\left(F_{3,n}-\frac{d_2F_{3,n+1}}{2\ell}-\frac{d_2F_{3,n-1}}{2\ell}\right),
\label{eq4veT}\\
V_{3,n}& \approx \frac{1}{4\pi\eta[-n\Omega]\ell }\left(\frac{F_{1,n}}{2}-\frac{d_1F_{1,n+1}}{8\ell}-\frac{d_1F_{1,n-1}}{8\ell}-\frac{d_2F_{1,n+1}}{8\ell}-\frac{d_2F_{1,n-1}}{8\ell}\right)\nonumber\\
&+\frac{1}{4\pi\eta[-n\Omega]\ell }\left(F_{2,n}-\frac{d_2F_{2,n+1}}{2\ell}-\frac{d_2F_{2,n-1}}{2\ell}\right)+\frac{F_{3,n}}{6\pi\eta[-n\Omega] a}.
\label{eq5veT}
\end{align}
Note that the couplings between different $n$-modes are involved in these equations.
Finally, substituting Eq.~(\ref{Fex}) into Eq.~(\ref{eq6}), we obtain
\begin{align}
\label{eq6veT}
&F_{1,n}+F_{2,n}+F_{3,n}=0.
\end{align}

The above set of equations constitute a matrix equation with infinite dimensions 
and cannot be solved in general.  
Under the assumption of $a\ll\ell$, however, we are allowed to consider only
$n=-1, 0, 1$ and further approximate as $F_{i,\pm2}\approx 0$. 
The justification of the latter approximation is also seen by solving 
Eqs.~(\ref{eq1veT}), (\ref{eq2veT}), (\ref{eq3veT}), (\ref{eq4veT}), 
(\ref{eq5veT}) and (\ref{eq6veT}) for $n=\pm2$ and taking the limit of $a\ll\ell$.
Hence the above set of equations can be solved for 18 unknowns, i.e., 
$V_{i,n}$ and $F_{i,n}$ for $i=1, 2, 3$ and $n=-1, 0, 1$.

The velocity of each sphere is simply obtained by the inverse Fourier transform,
$V_i(t)=(2\pi)^{{-1}} \int_{-\infty}^{\infty} {\rm d} \omega\, V_i(\omega) e^{i\omega t}$.
The average swimming velocity over one cycle of motion is then calculated by 
\begin{equation}
\label{barV}
\overline{V}=\frac{\Omega}{2\pi}\int_0^{2\pi/\Omega}{\rm d}t\, 
[V_1(t)+V_2(t)+V_3(t)]/3.
\end{equation}
Up to the lowest order terms in $a$, we finally obtain Eq.~(15).
In order to obtain more accurate higher order terms in $a$, one needs to 
take into account the higher order $n$-modes ($|n| \ge 2$).
Equations~(\ref{VEeq:res1}) and (\ref{VEeq:res4}) can be obtained similarly. 

\end{widetext}


\end{document}